\documentclass[12pt]{article}
\RequirePackage{lineno}  
\RequirePackage{xspace}
\usepackage[dvips]{color}
\usepackage{color}
\usepackage{epsfig}
\usepackage[tight]{subfigure}
\usepackage[figuresright]{rotating}
\usepackage{relsize}
\usepackage{minitoc}
\usepackage{amsmath}
\usepackage{wrapfig} 
\usepackage{amssymb}
\usepackage{graphicx,fancyhdr,amsmath,amsfonts,hangcaption,fancybox}

\newcommand\pubdate{\today}

\newcommand\pubnumber{}
\newcommand{\babar}{\mbox{\slshape B\kern-0.1em{\smaller A}\kern-0.1em B\kern-0.1em{\smaller A\kern-0.2em R} }}

\def\Title#1{\begin{center} {\Large #1 } \end{center}}
\def\Author#1{\begin{center}{ \sc #1} \end{center}}
\def\Address#1{\begin{center}{ \it #1} \end{center}}

\newcommand\pubblock{\rightline{\begin{tabular}{l} \pubnumber\\
         \pubdate  \end{tabular}}}
\newenvironment{Abstract}{\begin{center}{\bf Abstract}\end{center} \bigskip \begin{quotation}  }{\end{quotation}}
\newenvironment{Presented}{\begin{quotation} \begin{center} 
             PRESENTED AT\end{center}\bigskip 
      \begin{center}\begin{large}}{\end{large}\end{center} \end{quotation}}

\def\beq{\begin{equation}}
\def\eeq#1{\label{#1}\end{equation}}
\def\eeqn{\end{equation}}

\def\beqa{\begin{eqnarray}}
\def\eeqa#1{\label{#1}\end{eqnarray}}
\def\eeqan{\end{eqnarray}}

\let\bar=\overbar

\def\Dslash{\not{\hbox{\kern-4pt $D$}}}
\def\dslash{\not{\hbox{\kern-2pt $\del$}}}

\def\msb{{\bar{\ssstyle M \kern -1pt S}}}

\begin{document}
\begin{titlepage}
\pubblock

\vfill


\Title{Charmonium \& Bottomonium at $e^{+}e^{-}$ colliders}
\vfill
\Author{Valentina Santoro  \footnote{on behalf of the \babar Collaboration}}  
\Address{INFN Ferrara , via Saragat 1, Ferrara, 44122, Italy}
\vfill


\begin{Abstract}
We review recent charmonium and bottomonium spectroscopy results from the B-factories Belle and \babar. A particular focus is given to the new $\eta_{c}(1S)$ and $\eta_{c}(2S)$ measurement and to the radiative transitions from $\Upsilon (2S)$
 and $\Upsilon (3S)$ with converted photons.  

\end{Abstract}

\vfill

\begin{Presented}
The Ninth International Conference on\\
Flavor Physics and CP Violation\\
(FPCP 2011)\\
Maale Hachamisha, Israel,  May 23--27, 2011
\end{Presented}
\vfill

\end{titlepage}
\def\thefootnote{\fnsymbol{footnote}}
\setcounter{footnote}{0}
%


\section{Introduction}

In the last few years, quarkonium spectroscopy received significant contributions from the B-Factories \babar and Belle.
The impressive amount of data recorded by the two experiments
allowed to study rare decay chains and to look for undiscovered resonances
in the charmonium and bottomonium mass regions. The results presented here are based on different subsamples collected by the two experiments not only at the $\Upsilon(4S)$ but also at the $\Upsilon(3S)$ and $\Upsilon(2S)$ center mass energy.

\section{Charmonium Spectroscopy}

The charmonium spectrum consists of eight narrow states below the open charm threshold (3.73GeV) and several tens of states above the threshold, some of them wide (because they decay to $D\bar{D}$), some of them still narrow, because their decay to open charm is forbidden by some conservation rule. Below the threshold almost all states are well established, with the possible exception of the $1^{1}P_{1} ~(h_{c})$ which has been observed recently but whose properties still need to be measured accurately. On the other hand, very little is known above the threshold. Only one state has been positively identified as a charmonium $D$ state, the $\psi (3770)$, then there are several new ``Charmonium-like" states that are very difficult to accommodate in the charmonium spectrum. \\
The B-factories are an ideal place to study charmonium since charmonium states are copiously produced in a variety of processes: 
\begin{itemize}
\item B decays: B mesons decay to charmonium in about 3\% of cases.
\item ``Double-charmonium production": in this process, observed for the first time by Belle \cite{Abe:2002rb}, a $J/\psi$ or a $\psi(2S)$ is
produced together and exclusively with another charmonium state.
\item  $\gamma \gamma$ fusion: in this process 
two virtual photons are emitted by the colliding
$e^{+}e^{-}$ pair $(e^{+}e^{-}\to e^{+}e^{-} \gamma^{*} \gamma^{*}\to e^{+}e^{-} (c\bar{c})$). States with C=+1 are formed.
\item  Initial state radiation (ISR): where a photon
is emitted by the incoming electron or positron
$(e^{+}e^{-}\to \gamma (c\bar{c}))$, only  states with $J^{PC}=1^{--}$ are formed.
\end {itemize}

\subsection{$\eta_{c}(1S)$ \& $\eta_{c}(2S)$}

\begin{figure}[htb!]
\centering
\includegraphics[width=0.8\textwidth]{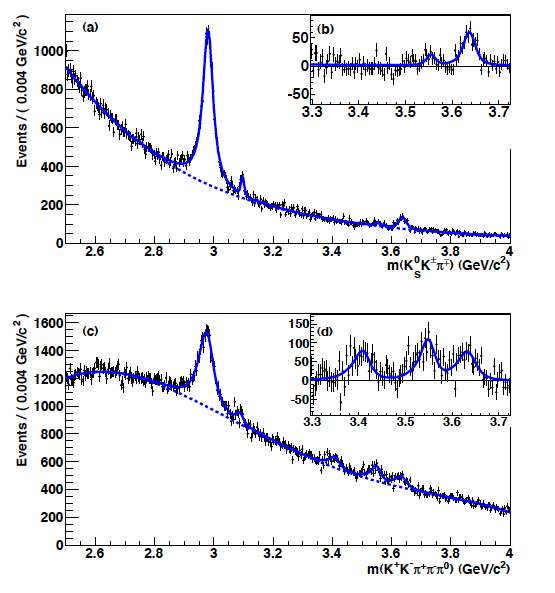}
\caption{Fit to the $K_{S}^{0}K^{\pm}\pi^{\mp}$ (a) and $K^{+}K^{-}\pi^{+}\pi^{-}\pi^{0}$ (c) mass spectra. The solid curves represent the total fit functions and the dashed curves show the combinatorial background contributions. The background-subtracted distributions are shown in (b) and (d).}
\label{fig:fig1}
\end{figure}

\begin{figure}[htb]
\centering
\includegraphics[width=0.6\textwidth]{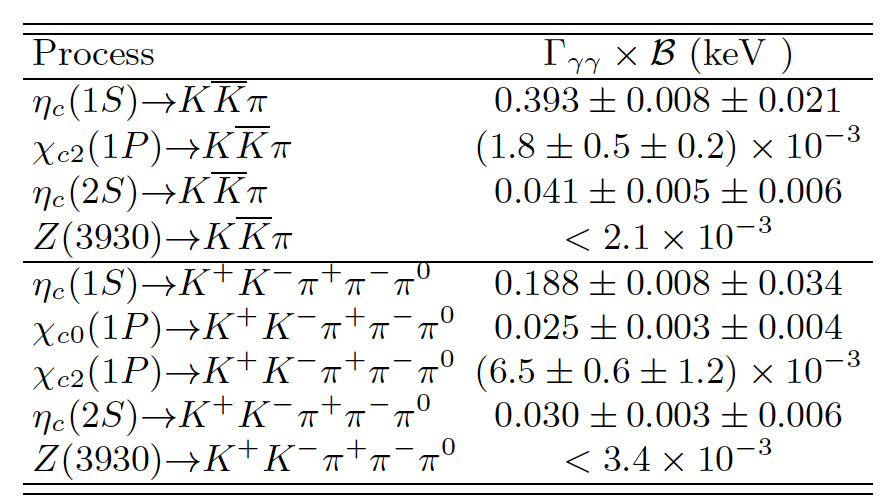}
\caption{Results for $\Gamma_{\gamma \gamma} \times {\cal B}$ for each resonaces in $K\bar{K}\pi$ and $K^{+}K^{-}\pi^{+}\pi^{-}\pi^{0}$ final states.  }
\label{fig:fig2}
\end{figure}

The $\eta_{c}(1S)$  was discovered by Crystal BALL in 1980  \cite{Partridge:1980vk} after that this resonance has been measured by several different experiments \cite{Aubert:2008kp,:2007vb,Asner:2003wv,Ambrogiani:2003md,Bai:2003et} with a large spread in the width measurements: in the case of $J/\psi$ and $\psi(2S)$ radiative decays its width is around 15 MeV while in B-decays and in $\gamma \gamma$ fusion is around 30MeV.
The $\eta_{c}(2S)$ was discovered by BELLE in 2002 \cite{Choi:2002na}
and until recently it has only been observed in exclusive decay to $K\bar{K}\pi$.
Precise measurement on $m(\eta_{c}(2S))$ will help discriminate among different charmonium models.\\
\babar studied using 519.2 $fb^{-1}$\cite{biassoni} collected at the $\Upsilon(4S)$, $\Upsilon(3S)$ and $\Upsilon(2S)$ the processes $e^{+}e^{-}\to e^{+}e^{-} \gamma \gamma$ $\to e^{+}e^{-} f$ where $f$ denotes $K^{+}K^{-}\pi^{+}\pi^{-}\pi^{0}$ or $K_{S}^{0}K^{\pm}\pi^{\mp}$ final states.  
 The allowed $J^{PC}$ values of the initial state are $0^{\pm+}, 2^{\pm+}, 4^{\pm+}, ...; 3^{++}, 5^{++}$.
Angular momentum, parity conservation, and charge
conjugation invariance, then imply that these quantum
numbers apply to the final states $f$ also, except that the
$K_{S}^{0}K^{\pm}\pi^{\mp}$ state cannot have $J^{P} = 0^{+}$. The $K_{S}^{0}K^{\pm}\pi^{\mp}$ and $K^{+}K^{-}\pi^{+}\pi^{-}\pi^{0}$ mass spectra are shown on Fig. \ref{fig:fig1}.
 There are signals at the position of the $\eta_{c}(1S)$, $J/\psi$, $\chi_{c0}(1P)$, $\chi_{c2}(1P)$ and $\eta_{c}(2S)$ states. The mass and the width of the $\eta_{c}(1S)$ and $\eta_{c}(2S)$, extracted using a binned extended maximum likelihood fit, are respectively $2982.5 \pm 0.4 \pm 1.4 ~\mathrm{MeV/c^{2}}$, $3638.5 \pm 1.5 \pm 0.8 ~\mathrm{MeV/c^{2}}$ and $32.1\pm  1.1 \pm 1.3 ~\mathrm{MeV}$, $13.4 \pm 4.6 \pm 3.2~\mathrm{MeV}$.  The results for $\Gamma_{\gamma\gamma} \times \cal{B}$  for each resonance in $K\bar{K}\pi$ and $K^{+}K^{-}\pi^{+}\pi^{-}\pi^{0}$ are reported in Fig. \ref{fig:fig2}. The first uncertainty is statistical, the second systematic. Upper limits are computed at 90\% confidence level.\\
 Another recent result on the $\eta_{c}(1S)$ and $\eta_{c}(2S)$ is the study of the $B^{\pm} \to K^{\pm} \eta_{c}(1S)$ and  $B^{\pm}\to K^{\pm}\eta_{c}(2S)$ performed by Belle \cite{belleetac}.
 Using 492 $fb^{-1}$ the $\eta_{c}(1S)$ and $\eta_{c}(2S)$ are reconstructed in the $K_{S}K\pi$ channel, in addition to these resonances is also found the presence of signal with the same final state of the resonant decay. This contribution is referred as the non-resonant contribution. Since the final state is the same, this amplitude interferes with the resonant component. The mass and the width for the $\eta_{c}(1S)$ are similar with and without interference but in the case of the $\eta_{c}(2S)$ the width for the interference hypothesis is: $6.6^{+8.4 +2.6} _{-5.1 -0.9 } \mathrm {MeV}$ while for the no interference hypothesis is $41.1 \pm 12.0^{+6.4}_{-10.9}~\mathrm{MeV}$. This result suggests that the interference effect is important.

\section{Bottomonium Spectroscopy}

The bound states of $b\bar{b}$, the bottomonium states, are the heaviest and most compact bound
states of quarks and anti quarks in nature. Spectroscopic measurements of fine and hyperfine structure splittings of hadronic and radiative
transitions in the bottomonium system are very important since they allow to test calculations of NRQCD, QCDME
and lattice QCD.\\ 
The B-factories are not usually considered  ideal facilities for the study of the bottomonium spectrum
since their energy is tuned to the peak of the $\Upsilon (4S)$ resonance, which decays in almost 100\% of cases to a $B\bar{B}$
pair. But both \babar and Belle have recorded data samples
at various energies in the bottomonium region that made possible to perform several measurements in the bottomonium sector and discoveries like the $\eta_{b}(1S)\cite{:2008vj}$, $h_{b}(1P)$\cite{veronique,bellehb}
 and $h_{b}(2P)$ \cite{bellehb}.

\section{Radiative transitions from the $\Upsilon(2S,3S)$ with converted photons}
\begin{figure}[htb!]
\centering
\includegraphics[width=0.6\textwidth]{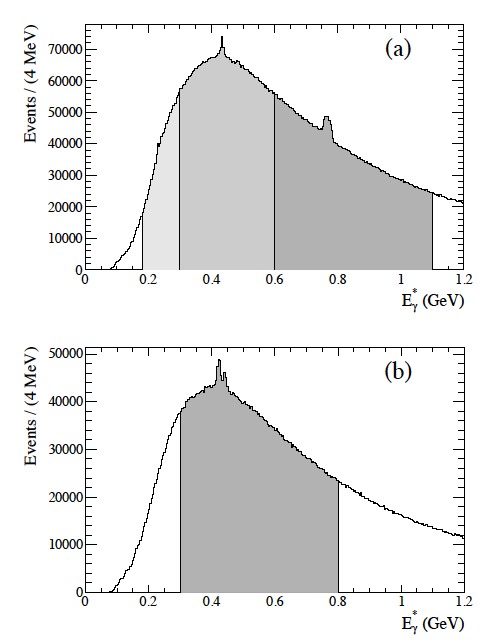}
\caption{Inclusive converted photon energy spectrum from (a) $\Upsilon(3S)$ and (b) $\Upsilon (2S)$ decays. The shaded areas indicate different regions of interest considered in details in this analysis.} 
\label{fig:fig3}
\end{figure}

\begin{figure}[htpb]
		\centering
			\subfigure[]{\includegraphics[width=7.5cm,height=4cm]{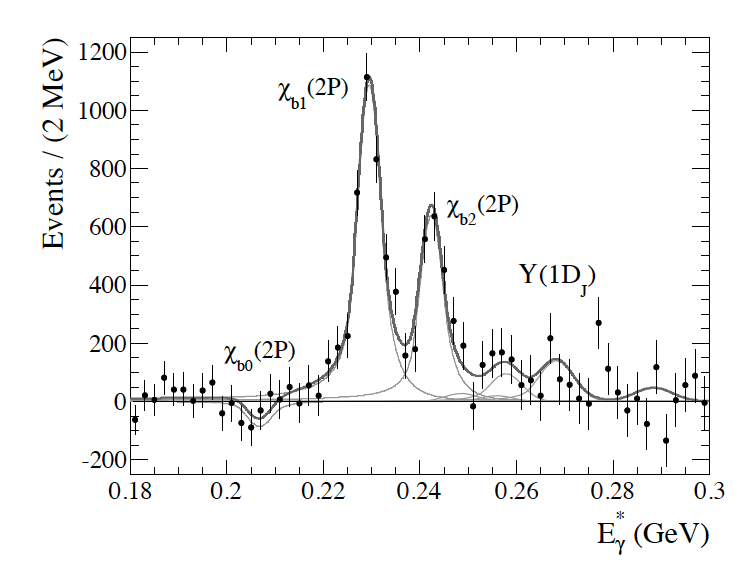}}
			\subfigure[]{\includegraphics[width=7.5cm,height=4cm]{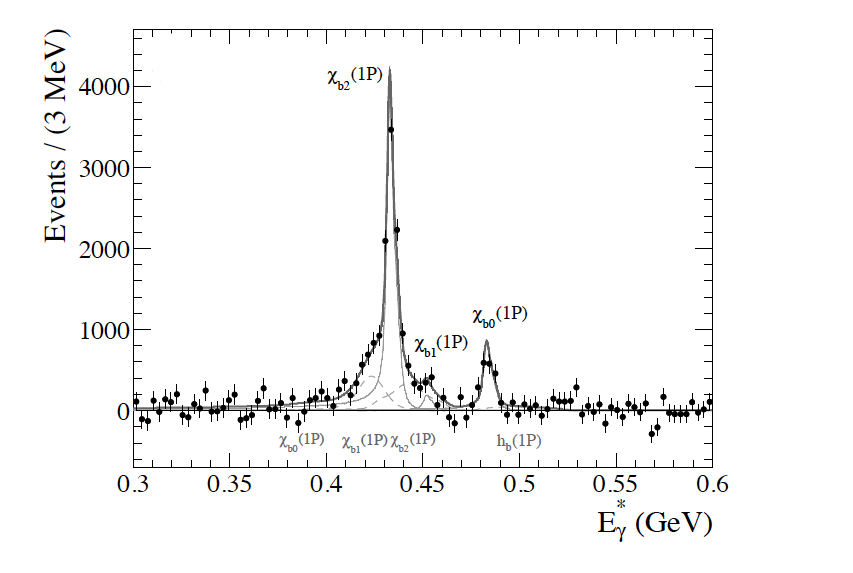}}
			\subfigure[]{\includegraphics[width=7.5cm,height=4cm]{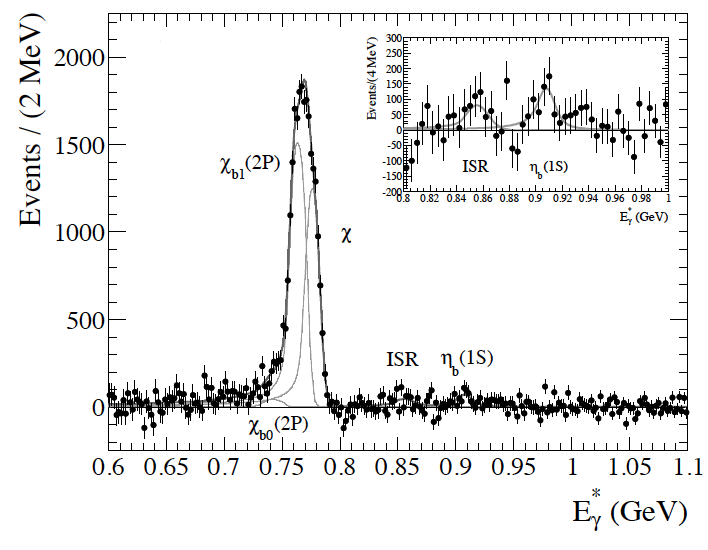}}
			\subfigure[]{\includegraphics[width=7.5cm,height=4cm]{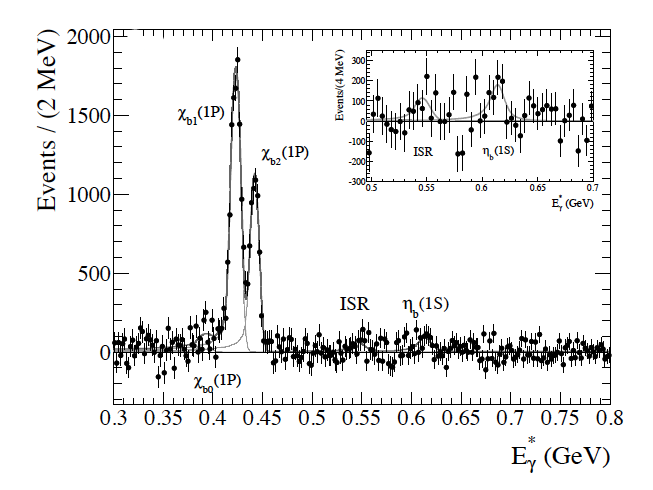}}
		\caption{Fit to the inclusive converted photon energy mass spectrum for different energies of the $\Upsilon(3S)$ and $\Upsilon(2S)$ data. }
	\label{fig:fig4}
\end{figure}

\babar studied the radiative transitions from the $\Upsilon(2S,3S)$ \cite{bryan} using photons converted to $e^{+}e^{-}$ pairs in the detector material, leading to a large improvement
in the $E^{*}_{\gamma}$ resolution at the cost of a reduced event yield. The signals appear as peaks in the inclusive photon energy spectra: $E(\gamma)_{CM}=((\sqrt{s})^{2}-m^{2}_{final~state})/2\sqrt{s})$. Fig. \ref{fig:fig3} shows the inclusive distributions of the resulting reconstructed converted photon energy. The data are divided into four energy ranges, as indicated by the shaded regions in Fig. \ref{fig:fig3}. These ranges and the corresponding bottomonium transitions of interest are, in $\Upsilon (3S)$ data: 
\begin{itemize}
\item $180 \leq E^{*}_{\gamma} \leq 300 \mathrm{MeV}$:  the main purpose of the fit to this region is to study the transitions $\chi_{bJ}(2P) \rightarrow \gamma \Upsilon (2S)$. The corresponding background subtracted plot is shown on Fig. \ref{fig:fig4}.(a). There are clear peaks for $\chi_{bJ}(2P)$ and even a hint for the $Y(1D)$.
\item $300 \leq E^{*}_{\gamma} \leq 600 \mathrm{MeV}:$  the main purpose of the fit to this region is to study the transitions $\Upsilon (3S) \rightarrow \gamma \chi_{bJ}(1P)$. The corresponding background subtracted plot is shown on Fig. \ref{fig:fig4}.(b). The clear peaks for $\chi_{bJ}(1P)$ are the most precise measurement for the transitions $\Upsilon(3S)\rightarrow \gamma \chi_{b0,2}(1P)$.
\item $600 \leq E^{*}_{\gamma} \leq 1100 \mathrm{MeV}:$ the main purpose of the fit to this region is to study the transitions $ ~\chi_{bJ}(2P)  \rightarrow \gamma \Upsilon (1S)$ and to look for $\Upsilon(3S) \to \gamma \eta_{b}(1S)$. The corresponding background subtracted plot is shown on Fig. \ref{fig:fig4}.(c). There are clear peaks for $\chi_{bJ}(2P)$, but there is no significance evidence for $\eta_{b}(1S)$.
\end{itemize}
and in the $\Upsilon(2S)$ data: 
\begin{itemize}
\item  $300 \leq E^{*}_{\gamma} \leq 800 \mathrm{MeV}:$ the main purpose of the fit to this region is to study the transitions $\chi_{bJ}(1P)\rightarrow \gamma \Upsilon(1S)$ and $\Upsilon(2S)\rightarrow \gamma \eta_{b}(1S)$.  The corresponding background subtracted plot is shown on Fig. \ref{fig:fig4}.(d). There are clear peaks for $\chi_{bJ}(1P)$, and a hint for the $\eta_{b}(1S)$. 
\end{itemize}

\begin{figure}[htb]
\centering
\includegraphics[width=0.6\textwidth]{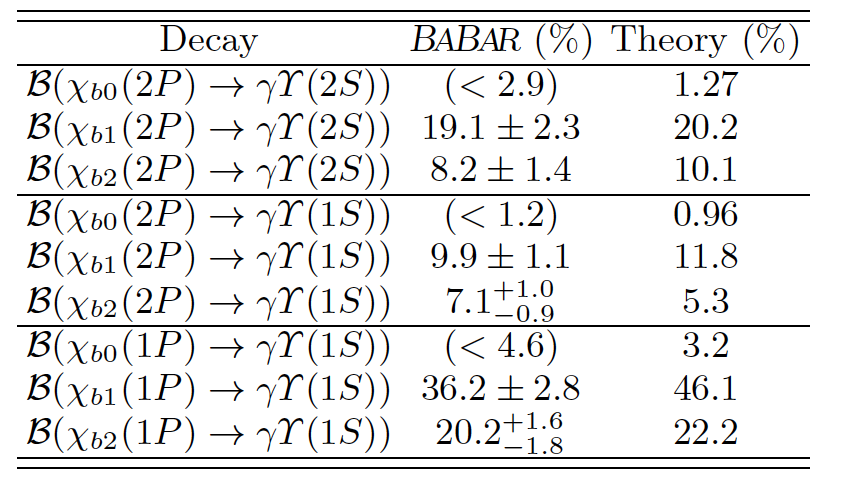}
\caption{Comparison of the experimental branching fraction from the \babar results and some theoretical predictions \cite{Kwong:1988ae}.}
\label{fig:fig5}
\end{figure}

The measured branching fractions with the comparison with the theoretical predictions are shown in Fig. \ref{fig:fig5}. There is a good overall agreement between theory \cite{Kwong:1988ae} and experiment.

\section{Inclusive dipion transition in $\Upsilon (3S)$ decays}
\begin{figure}[htb]
\centering
\includegraphics[width=0.8\textwidth]{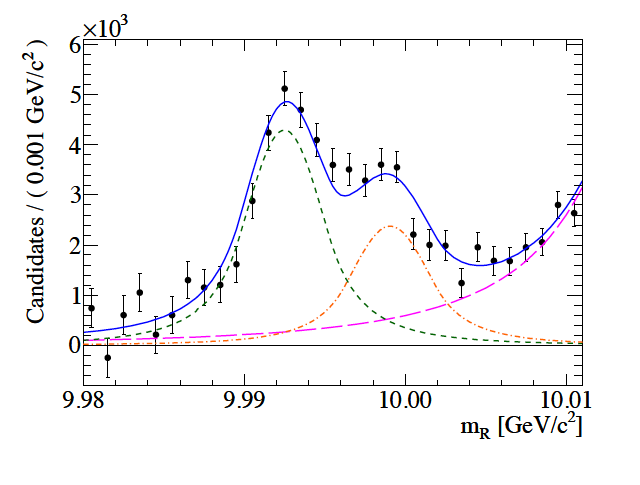}
\caption{The $m_{R}$ spectrum in the region between 9.98--10.01 GeV/c$^{2}$ after subtraction
of continuum and and $K^{0}_{S}$ background components: points represent data, while the curves represent the $\chi_{b1}(2P)\rightarrow \chi_{b1}(1P)$  (dashed),
$\chi_{b2}(2P)\rightarrow \chi_{b2}(1P)$ (dot-dashed), and $\Upsilon(3S) \to \Upsilon(2S)$ (long-dashed) components.  }
\label{fig:fig6}
\end{figure}

\babar in a search for the $h_{b}(1P)$ \cite{vincenzo} using a sample of $108 \times 10^{6} ~\Upsilon (3S)$ decays analyzed the inclusive 
 di-pion decays. These transitions are studied using a fit to the spectrum of the mass $m_{R}$ recoiling against the $\pi^{+}\pi^{-}$ system defined 
 as: $m_{R}^{2}=(M_{\Upsilon(3S)}-E^{*}_{\pi\pi})^{2}-|P^{*}_{\pi\pi}|^{2}$. The result showed no evidence for the
bottomonium spin-singlet state $h_{b}(1P)$ but the transitions $\chi_{bJ}(2P)\rightarrow \pi^{+}\pi^{-}\chi_{bJ}(1P), ~\Upsilon (3S)\rightarrow \pi^{+}\pi^{-}\Upsilon(2S)$, 
 and $\Upsilon (2S) \rightarrow   \pi^{+}\pi^{-}\Upsilon(1S)$ have been investigated. In Fig. \ref{fig:fig6} is shown the $m_{R}$ spectrum in the region between 9.98--10.01 GeV/c$^{2}$ after subtraction
of continuum and $K^{0}_{S}$ background components: there are clear signals for the  $\chi_{bJ}$ transitions, the extracted BF values are the following:
${\cal B}[\chi_{b1}(2P)\rightarrow \pi^{+}\pi^{-}\chi_{b1}(1P)]=9.2\pm0.6\pm 0.9\times 10^{-3}$ and  ${\cal B}[\chi_{b2}(2P)\rightarrow \pi^{+}\pi^{-}\chi_{b2}(1P)]=4.9\pm0.4\pm 0.6\times10^{-3}$.
In addition to these measurements is also extracted the ${ \cal B}[\Upsilon(3S) \rightarrow \pi^{+}\pi^{-}\Upsilon(2S)]=3.00 \pm 0.002 (stat) \pm 0.14(syst)\%$ and the difference between the mass of the $\Upsilon (3S)$ and $\Upsilon(2S)$ mass: $m(\Upsilon(3S)-\Upsilon(2S))=331.50 \pm 0.02(stat) \pm0.13(syst) \mathrm{MeV/c^{2}}$. The last two values are more precise that the current world average \cite{PDG}.  

\section{Conclusion}

Charmonium and Bottomonium spectroscopy has been revitalized by the B-factories discoveries made possible by their very large data sample. \babar and Belle using 1.6 $ab^{-1}$ discovered five missing pieces in the charmonium and bottomonium spectrum ($\eta_{c}(2S),$ $\chi_{c2}(2P)$, $\eta_{b}(1S),~h_{b}(1P),~h_{b}(2P))$, several unexpected new states, and performed many precise measurements. New exciting results are still coming from \babar and Belle, but given the fact that the next generation of B-factories being realized, even more exciting new results can be anticipated in the not-too-distant future.


\begin{thebibliography}{99}


\bibitem{Abe:2002rb}
  K.~Abe {\it et al.}  [Belle Collaboration],
  Phys.\ Rev.\ Lett.\  {\bf 89}, 142001 (2002)
 
  
\bibitem{Partridge:1980vk}
  R.~Partridge {\it et al.},
  Phys.\ Rev.\ Lett.\  {\bf 45}, 1150 (1980).

\bibitem{Aubert:2008kp}
  B.~Aubert {\it et al.}  [BABAR Collaboration],
  Phys.\ Rev.\  D {\bf 78}, 012006 (2008)
  [arXiv:0804.1208 [hep-ex]].

\bibitem{:2007vb}
  S.~Uehara {\it et al.}  [Belle Collaboration],
  Eur.\ Phys.\ J.\  C {\bf 53}, 1 (2008)
  [arXiv:0706.3955 [hep-ex]].

\bibitem{Asner:2003wv}
  D.~M.~Asner {\it et al.}  [CLEO Collaboration],
  Phys.\ Rev.\ Lett.\  {\bf 92}, 142001 (2004)
  [arXiv:hep-ex/0312058].

\bibitem{Ambrogiani:2003md}
  M.~Ambrogiani {\it et al.}  [Fermilab E835 Collaboration],
  Phys.\ Lett.\  B {\bf 566}, 45 (2003).
  
\bibitem{Bai:2003et}
  J.~Z.~Bai {\it et al.}  [BES Collaboration],
  Phys.\ Lett.\  B {\bf 555}, 174 (2003)
  [arXiv:hep-ex/0301004].


\bibitem{Choi:2002na}
  S.~K.~Choi {\it et al.}  [Belle collaboration],
  Phys.\ Rev.\ Lett.\  {\bf 89}, 102001 (2002)
  [Erratum-ibid.\  {\bf 89}, 129901 (2002)]
\bibitem{biassoni}
  P. del Amo Sanchez {\it et al.}  [\babar collaboration], arXiv:1103.3971v2
 
  \bibitem{belleetac}
  A. Vinokurova, {\it et al.}  [Belle collaboration], arXiv:1105.0978v2
  
\bibitem{:2008vj}
  B.~Aubert {\it et al.}  [BABAR Collaboration],
  Phys.\ Rev.\ Lett.\  {\bf 101}, 071801 (2008)
  [Erratum-ibid.\  {\bf 102}, 029901 (2009)]


\bibitem{veronique}
J. P. Lees {\it et al.} arXiv:1102.4565v1
\bibitem{bellehb}
I. Adachi {\it et al.} 	arXiv:1103.3419v1 

\bibitem{bryan}
 J. P. Lees {\it et al.} arXiv:1104.5254v1
 
\bibitem{Kwong:1988ae}
  W.~Kwong and J.~L.~Rosner,
  Phys.\ Rev.\  D {\bf 38}, 279 (1988).

\bibitem{vincenzo}
J. P. Lees {\it et al.} 	arXiv:1105.4234v1


\bibitem{PDG}
K. Nakamura  {\it et al.}, J. Phys. G {\bf 37}, 075021 (2010).
\end{thebibliography}
\end{document}